\documentclass[twocolumn,showpacs]{revtex4}
\usepackage{amssymb}
\usepackage{graphicx}
\usepackage{dcolumn}
\usepackage{bm}
\usepackage{amsmath}
\usepackage{epsfig}

\begin{document}

\title{Electric Deflection of Rotating Molecules}

\author{E. Gershnabel}
\author{I.Sh. Averbukh}
\affiliation{Department of Chemical Physics, The Weizmann Institute
of Science, Rehovot 76100, ISRAEL}
\begin{abstract}

We provide a theory of the deflection of polar and non-polar rotating molecules by inhomogeneous static electric field. Rainbow-like features in the angular distribution of the scattered molecules are analyzed in detail. Furthermore, we demonstrate that one may efficiently control the deflection process with the help of short and strong femtosecond laser pulses. In particular the deflection process may by turned-off by a proper excitation, and the angular dispersion of the deflected molecules can be substantially reduced. We study the problem both classically and quantum mechanically, taking into account the effects of strong deflecting field on the molecular rotations. In both treatments we arrive at the same conclusions. The suggested control scheme paves the way for many applications involving molecular focusing, guiding, and trapping by inhomogeneous fields.

\end{abstract}

\maketitle

\section{Introduction} \label{Introduction}

Deflection of molecules by inhomogeneous external fields is an important subject of molecular physics, which
continues to attract a lot of attention in recent years
\cite{McCarthy,Benichou,Loesch,Antoine,Deflection_general,Lens,Prism}.
The external fields can be magnetic or electric
\cite{McCarthy,Benichou,Loesch,Antoine}, or even optical fields of strong lasers
\cite{Deflection_general,Lens,Prism,Friedrich_ilya,Gordon_ilya,Fulton_ilya,Bishop_ilya,Seideman}.
By controlling molecular translational motion with external fields, novel elements of
molecular optics can be realized, including molecular lens
\cite{Deflection_general,Lens} and molecular prism \cite{Prism}.
Deflection by external fields is also used as a tool to measure molecular
polarizability \cite{Antoine} and molecular dipole moment. The mechanism of
molecular deflection by a nonuniform static electric field is rather
clear. For a non-polar molecule, the field induces  molecular
polarization, interacts with it, and deflects the molecules along
the interaction energy gradient. For a polar molecule, the field interacts with
the molecular permanent dipole moment as well. As most molecules
have anisotropic polarizability or/and a permanent dipole moment,
the deflecting force depends on the molecular orientation with
respect to the deflecting field. Previous studies on molecular
deflection have mostly considered randomly oriented molecules, for
which the deflection angle is somehow dispersed around the mean
value determined by the orientation-averaged polarizability (or
dipole moment). This dispersion was observed via broadening of the
scattered molecular beam, and was reported in the "two-wire" electric field
experiments \cite{Story,Antoine,Schafer} and also in the multipole electric
field experiments \cite{Brooks,Kramer,Lubbert}. More recently, this kind of rotation-induced
dispersion in molecular scattering by static electric fields was used as a selection tool in experiments
on laser-induced molecular alignment  \cite{post}. The
field-molecule interactions become intensity-dependent for strong
enough fields due to the field-induced modification of the molecular
angular motion \cite{Zon,Friedrich}. This adds a new ingredient for
controlling molecular trajectories  \cite{Seideman,Barker-new,Friedrich,Friedrich_ilya,Gordon_ilya}.

Recently, we showed that molecular deflection by strong fields of focused laser beams can be
significantly affected and controlled by \textit{pre-shaping}
molecular angular distribution \emph{before} the molecules enter the
interaction zone \cite{Gershnabel_Deflection}.  This can be done with the help of numerous recent
techniques for laser molecular alignment, which use single or
multiple short laser pulses (transform-limited, or shaped) to align
molecular axes along certain directions.  Short laser pulses excite
rotational wavepackets, which results in a considerable transient
molecular alignment after the laser pulse is over, i.e. at
field-free conditions (for reviews on field-free alignment,
see, e.g. \cite{Stapelfeldt,Stapelfeldt1}). Field-free alignment was
observed both for small diatomic molecules as well as for more
complex molecules, for which full three-dimensional control was
realized \cite{3D1,3D2,3D3}.

In the present paper we extend this approach to molecular deflection in static
electric fields, and  demonstrate that the average
scattering angle of deflected molecules and its distribution may be
dramatically modified by a proper field-free pre-alignment with ultra-fast lasers. An important
difference between the scattering in a static field and an optical field
\cite{Gershnabel_Deflection} is due to the role of the molecular permanent dipole moment.
Dipole interaction with the laser beams averages to zero because of the fast oscillations
of the optical fields, however this kind of interaction becomes dominant for polar molecules
placed in the static electric fields. In the present paper, we analyze in detail interaction of
rotating molecules with the static electric fields (taking into account both the dipole-type and polarization-type
interactions), and demonstrate that laser-induced pre-alignment provides a flexible tool for controlling
molecular motion in these fields.

In Sec. \ref{Deflection} we present the deflection scheme, and provide
heuristic arguments on the anticipated role of molecular rotation on the scattering
process (both for thermal molecules and molecules  pre-aligned by additional laser pulses).
In Sec. \ref{Classical} we provide a full classical treatment of the problem, which supports
the heuristic predictions of Sec. \ref{Deflection}. An alternative classical analysis based on the
formalism of adiabatic invariants is given in  Sec. \ref{Appendix}, and it leads to the same results.
In Sec. \ref{Quantum} we support our control approach by means of a full-scale quantum mechanical analysis.
Finally, we summarize our results in Sec. \ref{Conclusions}.

\section{Molecular Deflection} \label{Deflection}

\begin{figure}[htb]
\begin{center}
\includegraphics[width=8cm]{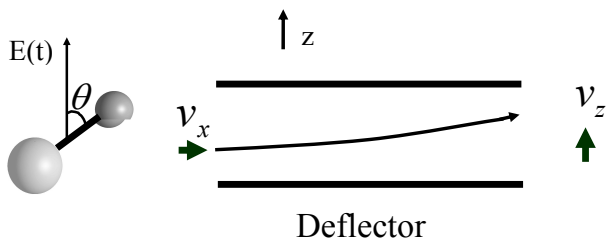}
\end{center}
\caption{The deflection scheme. Linear molecules, initially moving
in the $x$ direction (with velocity $v_x$), enter a static electric
field (directed along the $z$ axis). They  are deflected by the field
gradient, and get the deflection velocity $v_z$.} \label{QuantCS2P251}
\end{figure}

We consider a deflection scheme that is based on the interaction between
a linear molecule and an inhomogeneous static electric field. In particular,
we follow the lines of experiments \cite{Antoine,Tikhonov}, in which
a collimated particle beam goes through a long deflector
made of two cylindrical electrodes' faces.  Electrical field $F$ between
the two poles is equivalent to an "electrical two-wire field"
\cite{Ramsey}. This geometry  allows to obtain an
electric field $F$ and a field gradient $dF/dz$ which are nearly
constant over the width of the collimated molecular beam (see Fig. \ref{QuantCS2P251}).
\\

The interaction potential of a linear molecule in the static field
is given by:

\begin{equation}
U=-\frac{1}{2}F^2\left(\Delta\alpha
\cos^2\theta+\alpha_\perp\right)-\mu F
\cos\theta,\label{U_general_eq}
\end{equation}
where $F$ is the electric field; $\alpha_\parallel$ and
$\alpha_\perp$ are the components of the molecular polarizability
along the molecular axis, and perpendicular to it, respectively, and $\mu$ is the permanent dipole moment.
Here $\theta$ is the angle between the electric field
direction (along the laboratory $z$ axis) and the molecular axis. A
molecule initially moving along the $x$ direction   acquires a
velocity component $v_z$ along $z$-direction. We consider the
perturbation regime corresponding to a small deflection angle,
$\gamma\thickapprox v_z/v_x$ and, therefore, assume the molecules are
subject to the fixed values of the field and field gradient ($F$ and $\nabla F$,
respectively) inside the deflector.

The deflection velocity is given by:

\begin{equation}\label{Velocity_Deflection}
v_z
=-\frac{1}{M}\int_{-\infty}^{\infty}\left(\overrightarrow{\nabla}U\right)_z
dt,
\end{equation}
where $M$ is the mass of the molecules. The time-dependence of the
force (and of the potential $U$) in Eq.(\ref{Velocity_Deflection}) comes
from two sources: projectile motion of the molecule through the
deflector, and time variation of the angle
$\theta$ due to molecular rotation. For simplicity, we neglect the edge
effects at the entrance and exit of the deflector.

Since the rotational time scale is the shortest one in the problem,
we average the force over the fast rotation, and arrive at the
following expression for the deflection angle, $\gamma=v_x/v_z$:

\begin{equation}\label{Deflection_Angle}
\gamma=\left\{F\left[ \alpha_{||} {\cal A}_2 +
\alpha_{\perp}(1-{\cal A}_2)  \right]+\mu {\cal A}_1\right\}\frac{t_d\nabla F}{Mv_x}.
\end{equation}

Here ${\cal A}_{1,2}\equiv\overline{cos^{1,2}\theta}$ denotes the time-averaged
value of $\cos^{1,2}\theta$, and $t_d$ is the passage time through the deflector.
The quantities ${\cal A}_{1,2}$ depend on the
relative orientation of the vector of angular momentum and the
direction of the deflecting field. It is different for different
molecules of the incident ensemble, which leads to the randomization
of the deflection process.

\begin{figure}[htb]
\begin{center}
\includegraphics[width=2cm]{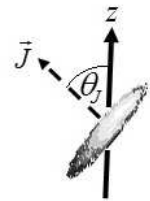}
\end{center}
\caption{A molecule rotates with a given angular momentum $\vec{J}$
that is randomly oriented in space. $\theta_J$ is the angle between
the angular momentum and the laboratory $z$ axis.}
\label{SimpleModel}
\end{figure}

We provide below some heuristic classical arguments on the
anticipated statistical properties  of ${\cal A}_{1,2}$ in the case of weak fields
that do not disturb significantly the molecular rotation.
We start with the simplest case of a linear molecule with $\mu=0$,  which rotates freely in a plane
perpendicular to the vector $\overrightarrow{J}$ of the angular
momentum (see Fig.(\ref{SimpleModel})).
\\

The projection of the molecular axis on the vertical $z$-direction
is given by:
\begin{equation}
\cos \theta(t)=  \cos(\omega t)
\sin\theta_J,\label{Simple_COS_model}
\end{equation}
where $\theta_J$ is the angle between $\vec{J}$ and $z$-axis, and
$\omega$ is the angular frequency of molecular rotation.

By averaging Eq. (\ref{Simple_COS_model}) over time, we obtain:

\begin{eqnarray}
{\cal A}_1&=&\overline{\cos\theta}=0\nonumber\\
{\cal A}_2&=&\overline{\cos^2\theta}=\frac{1}{2}
\sin^2\theta_J.\label{Simple_COS_avg_time}
\end{eqnarray}

For random and isotropic orientation of vector $\vec{J}$  in
space, the probability density for $\theta_J$ distribution is $1/2 \sin(\theta_J
)$.  The mean value of the deflection angle is then  $\langle
\gamma \rangle=\gamma_0$, where the constant $\gamma_0$ presents the
average deflection angle for an isotropic molecular ensemble:

\begin{equation}\gamma_0=\left[\alpha_{||}\frac{1}{3}+\alpha_{\perp}\frac{2}{3}\right]\frac{F\nabla
F t_d}{M v_x}.\label{gamma_isotropic}
\end{equation}

 Eq.(\ref{Simple_COS_avg_time}) allows us
to obtain the distribution function, $f({\cal A}_2)$ for ${\cal
A}_2$ (and the related deflection angle) from the known isotropic
distribution for $\theta_J$. Since the inverse function
$\theta_J({\cal A}_2)$ is multivalued, one obtains
\begin{equation}
f({\cal A}_2)=\sum_{i=1}^2 \frac{1}{2} \sin\theta_J^{(i)}{\left|
\frac{d{\cal
A}_2}{d\theta_J^{(i)}}\right|}^{-1}=\frac{1}{\sqrt{1-2{\cal
A}_2}},\label{Gamma_Dist}
\end{equation}
where we summed over the two branches of $\theta_J({\cal A}_2)$.
This formula predicts an \emph{unimodal  rainbow} singularity in the
distribution of the scattering angles at the maximal value
$\gamma=\gamma_0 (\alpha_{||} +\alpha_{\bot})/2\overline{\alpha}$ \
(for ${\cal A}_2=1/2$), and a flat step near the minimal one
$\gamma=\gamma_0 \alpha_{\bot}/\overline{\alpha}$ \ (for ${\cal
A}_2=0$). These results are similar to  the ones derived by us previously for
molecular scattering by oscillating optical fields \cite{Gershnabel_Deflection}.

As the next example, we consider the opposite case of polar molecules
with $\mu\neq 0$ and negligible polarization-type interaction.
For the sake of simplicity, we start with a $2D$ model, i.e. for a
molecule that rotates with no azimuthal momentum. In the limit of
$E/\mu F\ll 1$ ($E$ is the rotational energy), the molecular axis is
trapped by the electric field, and ${\cal A}_1\approx 1$. As $E$ is increased,
the molecules may still be trapped, but ${\cal A}_1<1$ and for
trapped molecules with high enough $E$, ${\cal A}_1<0$. The latter happens because the molecules
spend most of the time being against the electric field when performing nonlinear angular oscillations.
As the energy is increased even more, the molecules become untrapped, and
perform full rotations. In this case, we expect ${\cal A}_1<0$ due to the same reason: the molecules accelerate their rotation when the dipole moment tends to be parallel to the electric field, and they decelerate it when the dipole moment looks against the field. As a result, the time-averaged value of $\cos \theta$ is negative.
Considering a $2D$ molecular rotation in the presence of the electric field,  we can write:

\begin{equation}
dt=\sqrt{\frac{I}{2}}\frac{d\theta}{\sqrt{E+\mu
F\cos\theta}},\label{dt_vs_dtheta}
\end{equation}
where $I$ is the moment of inertia. Assuming the  untrapped regime ($\mu F/E<1$), the rotation period is given by \cite{Gradshteyn}:
\begin{eqnarray}
T_{period}&=&\sqrt{8 I}\int_{0}^{\pi}\frac{d\theta}{\sqrt{E+\mu
F\cos\theta}}\nonumber\\
&=& \sqrt{8 I}\frac{2}{\sqrt{E+\mu F}}F(\frac{\pi}{2},r)\nonumber\\\label{period_T_1}
\end{eqnarray}

The time averaged value of $\cos\theta$ is:

\begin{eqnarray}
{\cal A}_1&=&\frac{4}{T_{period}}\sqrt{\frac{I}{2}}\int_0^{\pi}\frac{d\theta}{\sqrt{E +\mu F\cos\theta}}\cos\theta\nonumber\\
&=&\left (\frac{E}{\mu F}+1\right ) \frac{E(\frac{\pi}{2},r)}{F(\frac{\pi}{2},r)}-\frac{a}{b}\label{cos_time_average}
\end{eqnarray}
Here $r\equiv\sqrt{\frac{2\mu F}{E+\mu F}}$;
$E\left(\frac{\pi}{2},r\right)$ and
$F\left(\frac{\pi}{2},r\right)$ are the first and second
order elliptic integrals, respectively. From Eq. (\ref{cos_time_average}) we learn that ${\cal A}_1\approx -\mu F/4E$ in the limit of weak fields, $\mu F/E\ll 1$.
This value is negative, as discussed above.

Summarizing, in the $2D$ approximation, ${\cal A}_1 = 1$ when $E/\mu F\ll 1$, it is close to $0$ when $\mu F/E\ll 1$, and it takes negative values in-between. This assumes the existence of a negative minimum of ${\cal A}_1$ as a function of $E/\mu F$.

The properties of the ${\cal A}_2$ quantity are somehow different in the considered $2D$ model.
For low $E$, the molecular rotation is suppressed, and ${\cal A}_2=1$. The period of angular oscillations of the trapped molecules ($\mu F/E>1$) is given by:

\begin{eqnarray}
T_{period}&=4 \sqrt{\frac{I}{2}}&\int_0^{\cos^{-1}\left(-\frac{E}{\mu F}\right)}\frac{d\theta}{\sqrt{E+\mu
F\cos\theta}}\nonumber\\
&=&\sqrt{\frac{I}{2}}\sqrt{\frac{2}{\mu
F}}F\left(\frac{\pi}{2},\frac{1}{r}\right)\label{period_T_2}
\end{eqnarray}

The time-averaged value of $\cos^2\theta$ is:

\begin{eqnarray}
{\cal
A}_2&=&\frac{4}{T_{period}}\sqrt{\frac{I}{2}}\int_0^{\cos^{-1}\left(-\frac{E}{\mu F}\right)}\frac{d\theta\cos^2\theta}{\sqrt{E+\mu
F\cos\theta}}\nonumber\\
&=&\frac{2}{3}\frac{E}{\mu F}+\frac{1}{3}-\frac{4}{3}\frac{E}{\mu
F}\frac{E\left(\frac{\pi}{2},\frac{1}{r}\right)}{F\left(\frac{\pi}{2},\frac{1}{r}\right)}
,\label{cos2_time_average}
\end{eqnarray}
where $r$, $E\left(\frac{\pi}{2},\frac{1}{r}\right)$ and
$F\left(\frac{\pi}{2},\frac{1}{r}\right)$ were defined above.

The function in Eq.(\ref{cos2_time_average}) has a local minimum at
${\cal A}_2=0.279$, which suggests a rainbow peak in the
distribution of ${\cal A}_2$ in the case of a smooth distribution of the parameter $E/\mu F$.
For high enough energy, the molecules   rotate almost as
free rotors, and we expect a rainbow peak in the ${\cal A}_2$
distribution at ${\cal A}_2=0.5$, as was suggested by Eq.(\ref{Gamma_Dist}).

Finally, in order to complete our analysis, we
consider the field-affected molecular rotation in $3D$ case, find numerically and plot
the quantities  of ${\cal A}_{1,2}$ (Figs. \ref{Classical_A_general
values} and \ref{Classical_A2_general_values}, respectively) for different values of dimensionless rotational energy and azimuthal canonical momentum (the details of the calculations can be found in the next section).
For the lowest possible negative values of the total energy, $E$ both plots demonstrate angular trapping  (${\cal A}_1=1, {\cal A}_2=1$). For small values of the azimuthal momentum we observe a negative shift (around $E\approx \mu F$) of the maximum of ${\cal A}_1$, the nature of which has been already discussed. In this limit, we also observe a minimum at
${\cal A}_2=0.274$ and a peak at ${\cal A}_2=0.5$, which is in agreement with the previous $2D$ model. For high
rotational energies (and high azimuthal momentum), ${\cal A}_1\approx 0$. Finally, for high azimuthal momentum
we observe also a strong decrease in the  ${\cal A}_2$ values, since  the molecules mainly
rotate in the $xy$ plane in this limit.

\begin{figure}[htb]
\begin{center}
\includegraphics[width=8cm]{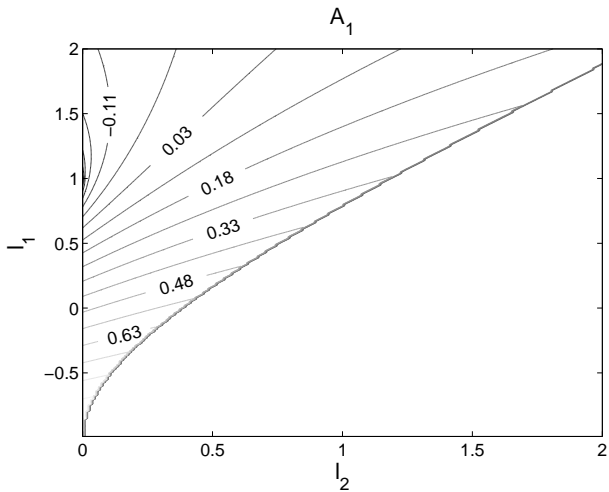}
\end{center}
\caption{Contour plot of ${\cal A}_1$ for different values of dimensionless rotational energy (i.e. $l_1\equiv\frac{E}{\mu F}$, vertical axis) and azimuthal canonical momentum (i.e. $l_2\equiv\frac{P_{\phi}^2}{2I\mu F}$, horizontal axis). The blank part of the figure corresponds to non-physical combinations of the parameter values. } \label{Classical_A_general values}
\end{figure}

\begin{figure}[htb]
\begin{center}
\includegraphics[width=8cm]{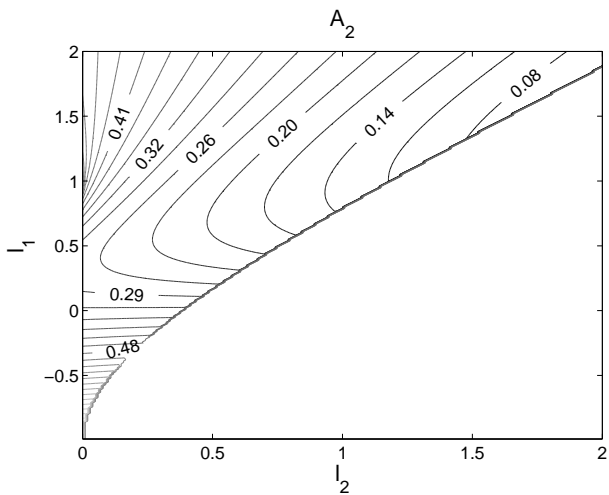}
\end{center}
\caption{Contour plot of ${\cal A}_2$ for different values of $l_1$ and $l_2$ ($l_{1,2}$ were defined in the caption of Fig. \ref{Classical_A_general values}). As in Fig. \ref{Classical_A_general values}, the blank part of the figure corresponds to non-physical combinations of the parameter values. For low azimuthal momentum, a minimum at ${\cal A}_2=0.274$ and a peak at ${\cal A}_2=0.5$ are observed, which suggests two rainbows in the ${\cal A}_2$ distribution.} \label{Classical_A2_general_values}
\end{figure}

\section{Classical Treatment} \label{Classical}
Consider a classical rigid rotor (linear molecule) described by the
Lagrangian:

\begin{eqnarray}\label{Lagrangian}
L&=&\frac{I}{2} \left(
\dot{\phi}^2\sin^2\theta+\dot{\theta}^2\right)\nonumber\\
&+&\frac{1}{2}F^2\left(\Delta\alpha\cos^2\theta+\alpha_{\bot}\right)\nonumber\\
&+&\mu F \cos\theta,
\end{eqnarray}
where $\theta$ and $\phi$ are Euler angles, and $I$ is the moment of
inertia. The canonical momentum for the $\phi$ angle
\begin{equation}
P_{\phi}= I \dot{\phi} \sin^2 \theta \label{Pphi}
\end{equation}
is a constant of motion as $\phi$ is a cyclic coordinate. The
canonical momentum $P_{\theta}$ is given by
\begin{equation}
P_{\theta}= I \dot{\theta}.  \label{Ptheta}
\end{equation}
The Euler-Lagrange equation for the $\theta$ variable is
\begin{equation}
\frac{d}{dt}\frac{\partial L}{\partial\dot{\theta}}-\frac{\partial
L}{\partial\theta}=0, \label{EL}
\end{equation}
which leads to
\begin{equation}
\frac{d^2\theta}{dt^2}=
\left(\frac{P_{\phi}}{I}\right)^2\frac{\cos\theta}{\sin^3\theta}-\frac{F^2\Delta\alpha}{I}\sin\theta\cos\theta-\frac{\mu
F}{I}\sin\theta.\label{thetaeq}
\end{equation}
When considering a thermal ensemble of molecules, it is convenient
to switch to dimensionless variables, in which the canonical momenta
are measured in the units of $p_{th}=I\omega_{th}$,  with
$\omega_{th}=\sqrt{k_B T/I}$, where $T$ is the temperature
\cite{Gershnabel}, and $k_B$ is the Boltzmann's constant. By setting $P_{\phi}'=P_{\phi}/p_{th}$,
$P_{\theta}'=P_{\theta}/p_{th}$, and $t'=\omega_{th}t$, Eq.(\ref{thetaeq}) becomes:
\begin{equation}
\frac{d^2\theta}{dt'^2}=P_{\phi}'\frac{\cos\theta}{\sin^3\theta}-C\sin\theta\cos\theta-D\sin\theta
\label{thetaeq_dimensionless}
\end{equation}
where $C\equiv F^2\Delta\alpha/(k_BT)$, and $D\equiv \mu F/(k_BT)$.

Considering a deflecting field that is adiabatically increasing to
its final value $F$ (adiabatic with respect to the molecular
rotational dynamics), we numerically solve Eq.(\ref{thetaeq_dimensionless}) and find the time dependent values of
$\cos\theta(t)$ and $\cos^2\theta(t)$.

In order to find the ${\cal{A}}_{1,2}$, we calculate:
\begin{equation}
{\cal{A}}_{1,2}(t)=\frac{1}{t-t_F}\int_{t_F}^{t}\cos^{1,2}\theta
dt ,\label{average_time_align}
\end{equation}
and consider ${\cal{A}}_{1,2}=\overline{\cos^{1,2}\theta}$ as the limit value to which  ${\cal{A}}_{1,2}(t)$
converges as $t-t_F\rightarrow \infty$. Here $t_F$ is the rising time
in which the deflecting field reaches its maximal value $F$.

The probability distribution of ${\cal{A}}_{1,2}$ is given by:
\begin{eqnarray}
f({\cal A}_{1,2})&=&\int\int\int\int d\theta(0) d\phi(0)
dP_{\theta}'(0) dP_{\phi}'(0)
\nonumber\\
&\times&\delta({\cal A}_{1,2}-\overline{\cos^{1,2}\theta})\nonumber\\
&\times&f(\theta(0),\phi(0),P_{\theta}'(0),P_{\phi}'(0)), \label{Accurate_Dist}
\end{eqnarray}
where
\begin{equation}
f=\frac{1}{8\pi^2}\exp\left[-\frac{1}{2}\left(P_\theta'^2+\frac{P_{\varphi}'^2}{\sin^2\theta}\right)\right]
\label{BoltDist}
\end{equation}
is the thermal distribution function.

\begin{figure}[htb]
\begin{center}
\includegraphics[width=8cm]{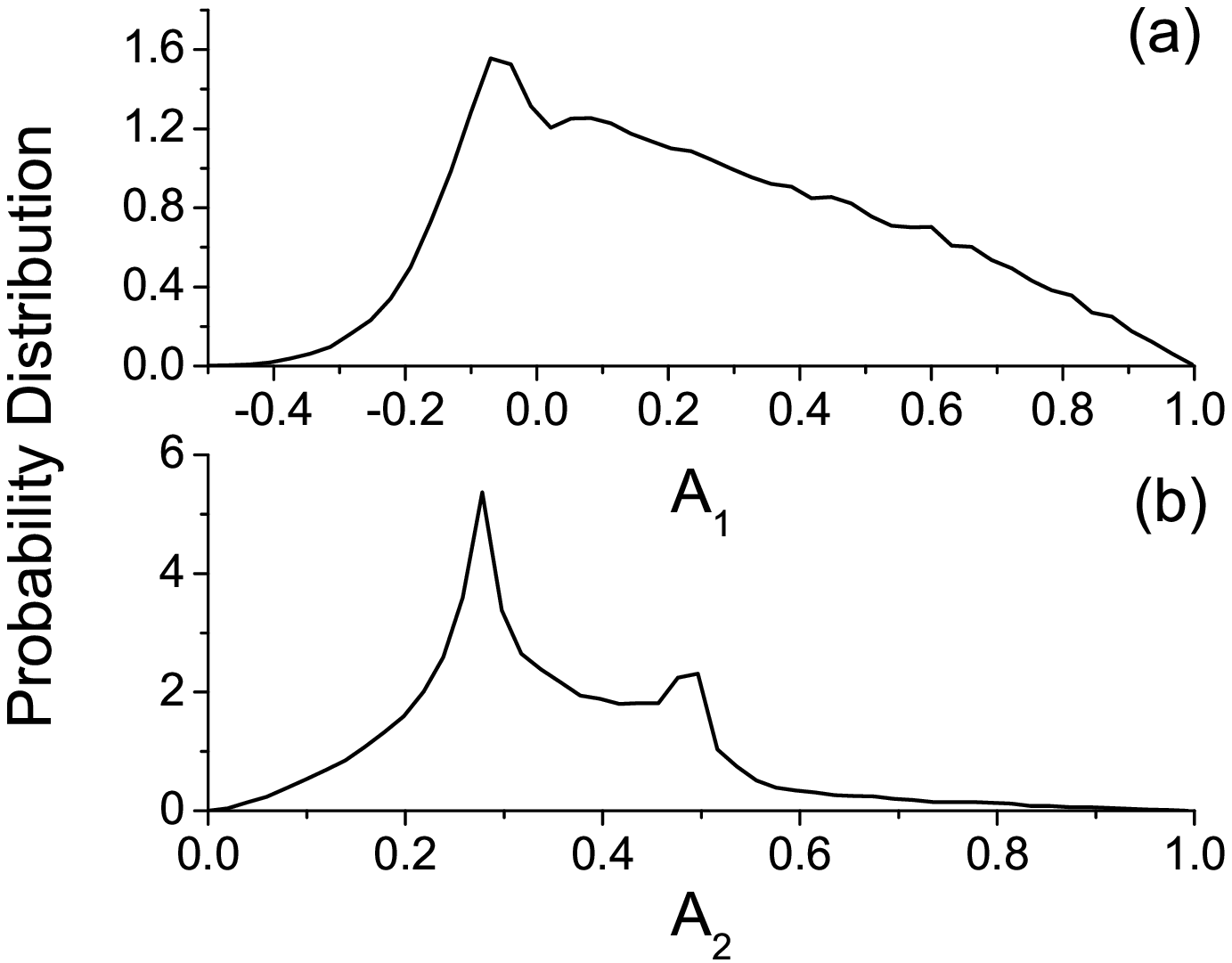}
\end{center}
\caption{Statistical distributions of (a) ${\cal A}_{1}$ and (b) ${\cal A}_{2}$ for a thermal beam of $KCl$ molecules in a moderate deflecting field  ($1.8\cdot 10^6V/m$). $J_T=5$; $C=5.81\cdot 10^{-6}$ ; $D=0.96$  } \label{Classical_Weak_No_Pre}
\end{figure}

\begin{figure}[htb]
\begin{center}
\includegraphics[width=8cm]{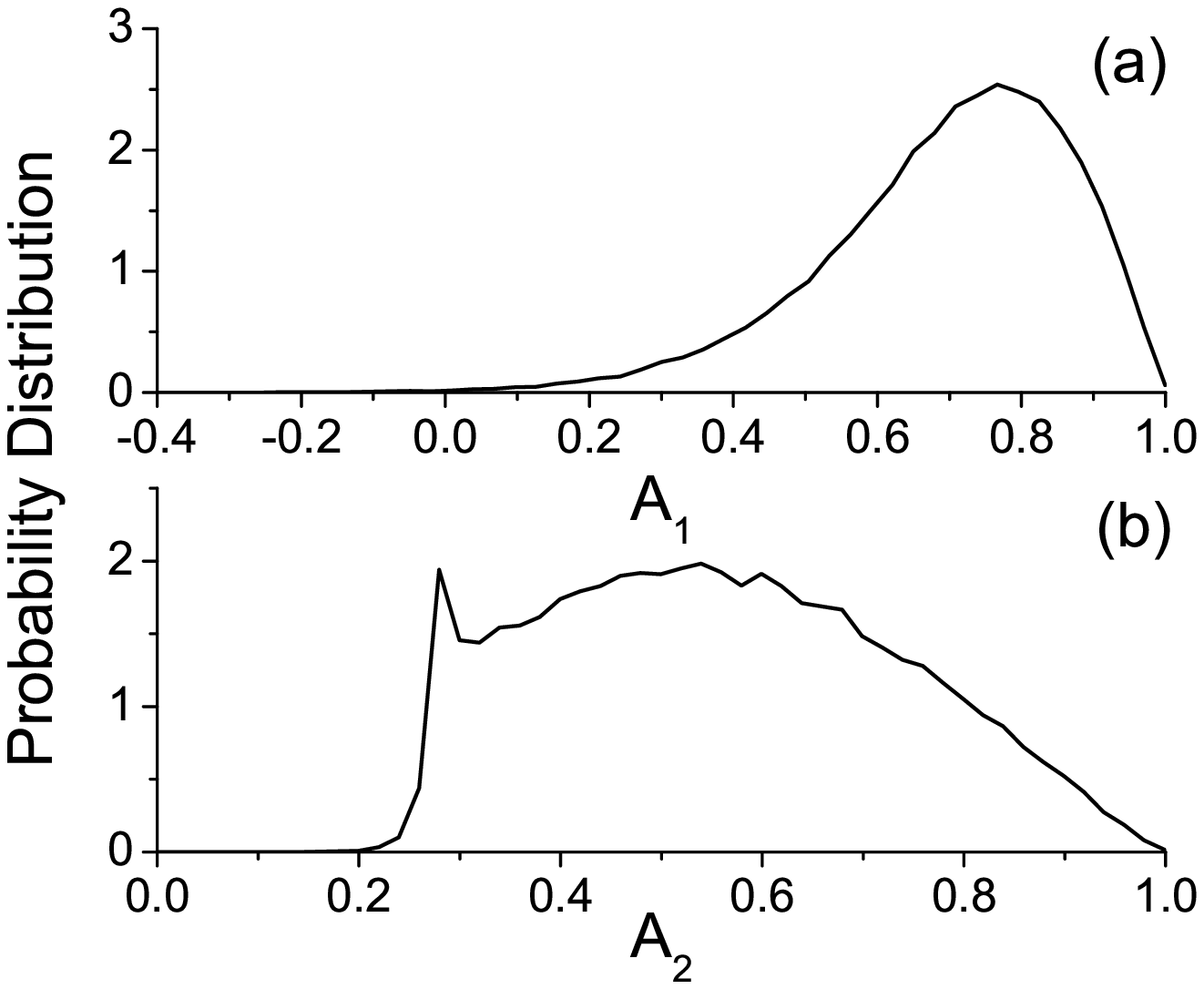}
\end{center}
\caption{Statistical distributions of (a) ${\cal A}_{1}$ and (b) ${\cal A}_{2}$ for a thermal beam of $KCl$ molecules in a strong deflecting field  ($1.8\cdot 10^7V/m$). $J_T=5$; $C=5.81\cdot 10^{-4}$ ; $D=9.62$.} \label{Classical_Normal_No_Pre}
\end{figure}

{\bf Deflection of thermal molecules.}  Using this approach, we considered distribution functions for ${\cal A}_{1,2}$ (and corresponding distributions  of the deflection angle $\gamma$) for a thermal beam of  $KCl$ molecules. For the chosen rotational temperature $T=4.63K$, the typical "thermal" value of the angular momentum is $J_T=5$, where $J_T=\sqrt{k_B T/(hB_rc)}$, $B_r$ is the rotational constant and $c$ is the speed of light.
We plot the  distribution functions for moderate ($1.8\cdot 10^6V/m$) and strong ($1.8\cdot 10^7V/m$) values of the deflecting field at Figs. \ref{Classical_Weak_No_Pre} and
\ref{Classical_Normal_No_Pre}, respectively.
For the  moderate field, $C=5.81\cdot 10^{-6}$ and $D=0.96$, so that
the dipole-field interaction is comparable with the
typical thermal rotational energy, while the polarization-type interaction is negligible.
In this case, a sizable portion of molecules are trapped by the field, which is reflected
in the high positive values of ${\cal A}_1$ (Fig.
\ref{Classical_Weak_No_Pre}a). The untrapped molecules are performing full rotations, and
they contribute to the negative values of ${\cal A}_1$. Fig. \ref{Classical_Weak_No_Pre}b presents the
distribution of ${\cal A}_2$, and it shows two rainbows (the first one at approximately $0.28$,
and the second one at $0.5$), as is expected from the discussion
in the previous section. The distribution of the deflection angles  (not shown here) is
similar to the distribution of ${\cal A}_1$, as the contribution from the polarization-type interaction (proportional to ${\cal A}_2$) is negligible in the considered numerical example. The ${\cal A}_2$ distribution for the $KCl$ molecule may be directly measured in a deflection experiment that combines a homogeneous static field and an inhomogeneous laser field. The static field will define the distributins of ${\cal A}_1$ and ${\cal A}_2$, and the laser field will deflect the molecules according to the ${\cal A}_2$ values \cite{Gershnabel_Deflection}.

In the case of a strong field ($1.8\cdot 10^7V/m$) shown at Fig. \ref{Classical_Normal_No_Pre},
$D=9.62$, and $C$ is still negligible. Now the dominant portion of molecules
is highly trapped by the electric field, and
the distribution of ${\cal A}_1$  is shifted accordingly to the higher
positive values (Fig. \ref{Classical_Normal_No_Pre}a). The rainbow
at $0.5$ in the ${\cal A}_2$ distribution (Fig. \ref{Classical_Normal_No_Pre}b) practically disappears due to the
increased amount of molecules with the suppressed rotation.

\begin{figure}[htb]
\begin{center}
\includegraphics[width=8cm]{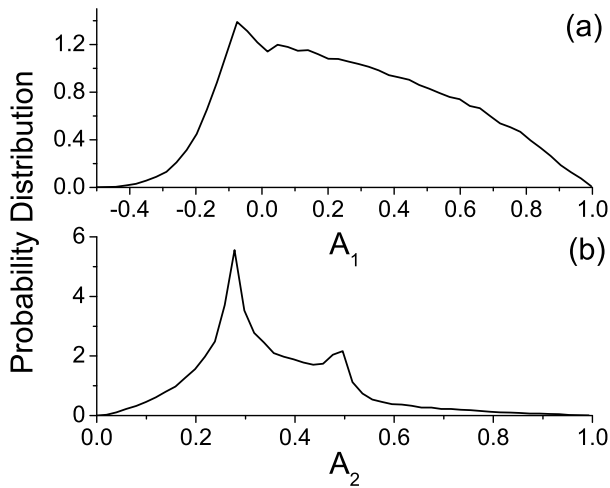}
\end{center}
\caption{Statistical distributions of (a) ${\cal A}_{1}$ and (b) ${\cal A}_{2}$ for a thermal beam of $KCl$ molecules at higher temperature that corresponds to $J_T=15$. Though the field is strong: $1.8\cdot 10^7V/m$ ($C=6.46\cdot 10^{-5}$ ; $D=1.07$), $D$ value is similar to the one of Fig. \ref{Classical_Weak_No_Pre} and the curves are correspondingly similar. } \label{ClassicalJT15}
\end{figure}

Finally, in Fig. \ref{ClassicalJT15} we present the case of a strong deflecting field, but at higher temperature. Though the field is as strong as in Fig. \ref{Classical_Normal_No_Pre}, $D$ value is similar to the one of Fig. \ref{Classical_Weak_No_Pre}, and the curves are correspondingly similar.

\begin{figure}[htb]
\begin{center}
\includegraphics[width=8cm]{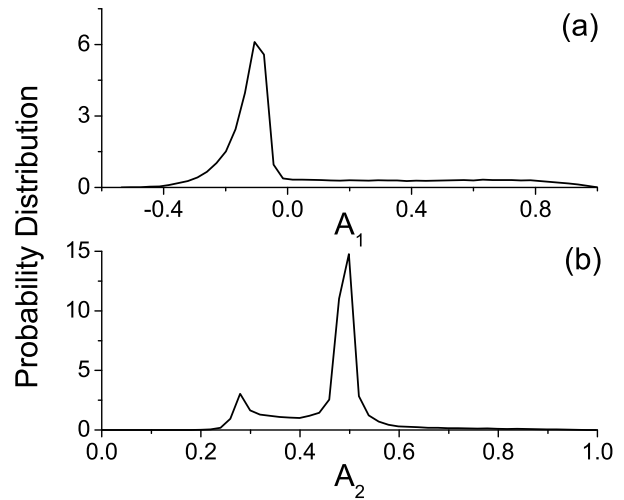}
\end{center}
\caption{The statistical distributions of (a) ${\cal A}_{1}$ and (b) ${\cal A}_{2}$ for $prealigned$ $KCl$ molecules ($P=25$). The prealigning pulse is parallel to the deflecting field (along $z$-axis). The deflection field is strong: $1.8\cdot 10^7V/m$, and $J_T=5$ ($C=5.81\cdot 10^{-4}$ ; $D=9.62$).}
\label{Classical_Parallel}
\end{figure}

{\bf Deflection of pre-aligned molecules.}
Assume now that the molecules are subject to a femtosecond pre-aligning
pulse polarized in $z$-direction at $t=0$, before they enter the deflecting field.
The  interaction of molecular permanent dipole moment with the laser pulse averages to zero because of the fast optical oscillations. The polarization-type interaction is given by the first term in Eq. (\ref{U_general_eq}),
in which $F$ is replaced by the envelope $\epsilon$ of the femtosecond pulse, and an additional factor of $0.5$ is added due to the oscillatory nature of the optical field.  We
assume  that  the  pulse is short compared to the rotational
period of the molecules, and consider it as a delta-pulse. The
rotational dynamics of the laser-kicked molecules is then described
by the same formalism as above, but with $P_{\theta}' (0)$
replaced by
\begin{equation}
 P_{\theta}'(0)\rightarrow P_{\theta}'(0)-P_s'\sin(2\theta(0)).
\end{equation}
Here $P_s '=P\hbar /\sqrt{k_BTI}$ is a properly normalized kick
strength of the laser pulse, with $P$ given by:
\begin{equation}
P=\left(1/4\hbar\right)\cdot
(\alpha_{||}-\alpha_{\bot})\int_{-\infty}^{\infty}\epsilon^2(t)dt.\label{Kick_P}
\end{equation}
Here we assumed the vertical polarization (along $z$-axis) of the
pulse. Physically, the dimensionless kick strength, $P$ equals to
the typical amount of angular momentum (in the units of $\hbar$)
supplied by the pulse to the molecule. For example, in the case of $KCl$ molecules, $P=25$ corresponds to the excitation by $2ps$ (FWHM) laser pulses with the maximal intensity of $5.8\cdot 10^{12}W/cm^2$. The distribution functions for kicked molecules are shown in
Fig. \ref{Classical_Parallel}. The kick parallel to the deflecting field
increases the rotational energy of the molecules (i.e.
makes them untrapped), while keeping unchanged the value of the azimuthal momentum.
As was explained in the previous section, untrapped molecules with relatively low azimuthal momentum
contribute to the negative shift of the peak in the ${\cal A}_1$ distribution function
(Fig. \ref{Classical_Parallel}a). Since most of the molecules became
untrapped, the rainbow around $0.5$ in the distribution of ${\cal
A}_2$ becomes the dominant rainbow. For the numerical example under consideration, the distribution of the deflection angles has the same shape as the distribution for ${\cal A}_1$ (with a proper scaling). As follows from Fig. \ref{Classical_Parallel}a , a prealigning laser pulse applied parallel to the direction of the deflecting field leads to a dramatic narrowing in the distribution of the scattering angles, and increases the brightness of the molecular beam deflected by a static electric field.

In the case of an aligning pulse in the $x$ direction (perpendicular to the deflecting field), both
$P_{\theta}'(0)$ and $P_{\phi}'(0)$ are replaced by:
\begin{eqnarray}
P_{\theta}'(0)&\rightarrow& P_{\theta}'(0)+P_s'\cos^2\phi(0)\sin(2\theta(0))\nonumber\\
P_{\phi}'(0)&\rightarrow&
P_{\phi}'(0)-P_s'\sin^2(\theta(0))\sin(2\phi(0))\label{PerpendicularPulses}
\end{eqnarray}

\begin{figure}[htb]
\begin{center}
\includegraphics[width=8cm]{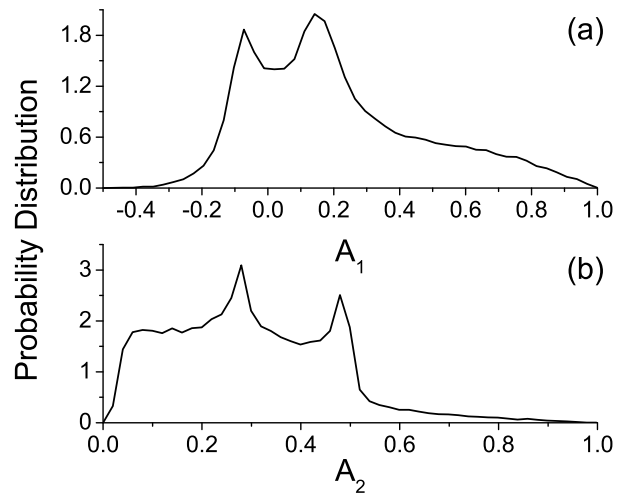}
\end{center}
\caption{The distributions of (a) ${\cal A}_{1}$ and (b) ${\cal A}_{2}$ for $prealigned$ $KCl$ molecules. The conditions are similar to those of Fig. \ref{Classical_Parallel}, except the polarization of the prealignment pulse, which is perpendicular to the deflecting field.}
\label{Classical_Perpendicular}
\end{figure}

The distribution functions for ${\cal A}_{1,2}$ in this case are shown in Fig. \ref{Classical_Perpendicular}.
Such a pulse forces the molecules to rotate preferentially in the planes containing the $x$ axis. In a previous
work \cite{Gershnabel_Deflection}, we showed that the distribution of
${\cal A}_2$ has two rainbows at $0$ and $0.5$ in the absence of the static field. In the present case,
as can be seen from Fig. \ref{Classical_Perpendicular}b, the distribution of ${\cal A}_2$ still preserves the rainbow at $0.5$, but the rainbow at $0$ is
smeared due to the effect of the strong deflecting field ($D=9.62$). Nevertheless, we still may
observe a considerable concentration of molecules at low ${\cal A}_2$ values.
The peak at  ${\cal A}_2 \approx 0.28$ is still present  due to the reasons explained in the previous section.
The two rainbows  are  due to molecules that rotate with
small azimuthal momentum. The distribution of ${\cal A}_1$ (Fig.
\ref{Classical_Perpendicular}a) has now a strong peak at low
positive ${\cal A}_1$ values, since a great portion of the molecules
is  freely rotating with large azimuthal momentum provided by the laser kick.

Finally, before proceeding to the quantum treatment of the same problem, we refer the reader to
Sec. \ref{Appendix}, in which an alternative approach to the classical
calculation of ${\cal A}_{1,2}$ is given by  means of the formalism of adiabatic
invariants.

\section{Quantum Treatment} \label{Quantum}

For a more quantitative treatment, involving analysis of the
relative role of the quantum and thermal effects on one hand, and
the strength of the pre-aligning pulses on the other hand, we
consider quantum-mechanically the deflection of a linear molecule
described by the Hamiltonian:

\begin{equation}
\hat{H}=\frac{\hat{J}^2}{2I}-\mu F
\cos\theta-\frac{1}{2}F\left[(\alpha_{||}-\alpha_{\bot})\cos^2\theta+\alpha_{\bot}\right],\label{Deflection_Hamiltonian}
\end{equation}
where $\hat{J}$ is the
angular momentum operator.

Without the electric field $F$, the eigenfunctions of the free-space
molecule are given by the free-rotor eigenfunctions $|J,m\rangle$. Before the molecules enter the deflecting field, we prealign them by a short femtosecond laser pulse. Such a pulse creates a rotational wave packet of the $|J,m\rangle$ states. After the prealigning laser pulse is over, the molecules enter adiabatically the region of the static field. Then, each $|J,m\rangle$ state (within the wave packet) transforms into the corresponding eigenstate $|\bar{J},m\rangle$, where $\bar{J}$ is the quantum number associated adiabatically with the quantum number $J$. The relation between
$|\bar{J},m\rangle$ and the free-rotor eigenfunctions may be described by:
\begin{equation}
|\bar{J},m\rangle=\sum_{J=|m|}^{\infty}\beta_{J,m}^{\bar{J}}|J,m\rangle\label{Free_Bounded_Relation}
\end{equation}
The force, ${\cal F}$ acting on the molecule is given by:
\begin{equation}
{\cal F}=-\nabla E=-\frac{\partial E}{\partial
F}\frac{dF}{d z}.\label{Force_quantum}
\end{equation}
The first derivative is obtained by the means of the Hellman-Feyman
theorem, that is being in an eigenstate $|\overline{J},m\rangle$,
\begin{equation}
\frac{\partial E_{\overline{J},m}(F)}{\partial
F}=\langle \overline{J},m|\frac{\partial H}{\partial
F}|\overline{J},m\rangle. \label{Hellman_Feyman}
\end{equation}

From Eqs. (\ref{Deflection_Hamiltonian}), (\ref{Force_quantum}) and (\ref{Hellman_Feyman}) it is clear that the deflection angle of a molecule in a $|\overline{J},m\rangle$ state is given by Eq. (\ref{Deflection_Angle}), in which ${\cal A}_{1,2}$ are replaced by:

\begin{eqnarray}
{\cal A}_1^{\overline{J},m}&=&\langle
\overline{J},m|\cos\theta|\overline{J},m\rangle\nonumber\\ {\cal
A}_2^{\overline{J},m}&=&\langle
\overline{J},m|\cos^2\theta|\overline{J},m\rangle \label{Quantum_A}
\end{eqnarray}

\begin{figure}[htb]
\begin{center}
\includegraphics[width=8cm]{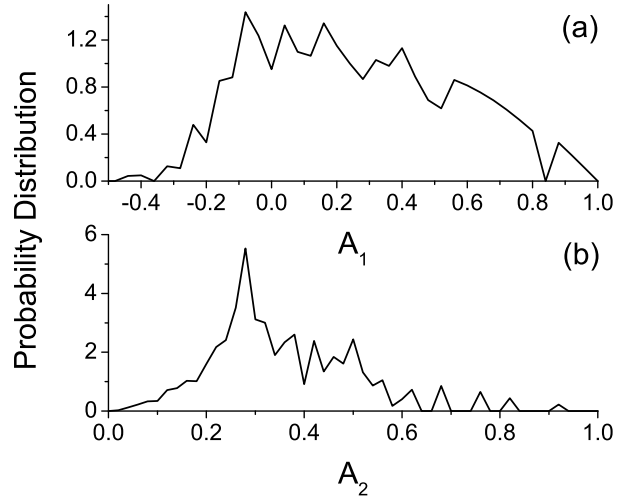}
\end{center}
\caption{Quantum distributions of (a) ${\cal A}_1$ and (b) ${\cal A}_2$ for a thermal beam of $KCl$ molecules in a strong field ($1.8\cdot 10^7V/m$) and $J_T=15$ ($C=6.46\cdot 10^{-5}$ ; $D=1.07$). These graphs are similar to the classical graphs of Fig. \ref{ClassicalJT15}.} \label{QuantumThermal}
\end{figure}

In the quantum case, the continuous distribution of the angles
$\gamma$ is replaced by a set of discrete lines, each of them
weighted by the thermal population of the state $|J,m\rangle$. Fig.
\ref{QuantumThermal} shows the distributions of ${\cal
A}_{1,2}^{\overline{J},m}$ in the thermal case (i.e. without prealignment). These results can be compared to their classical analogs in
Fig. \ref{ClassicalJT15}, where the same structure is
seen.

\begin{figure}[htb]
\begin{center}
\includegraphics[width=8cm]{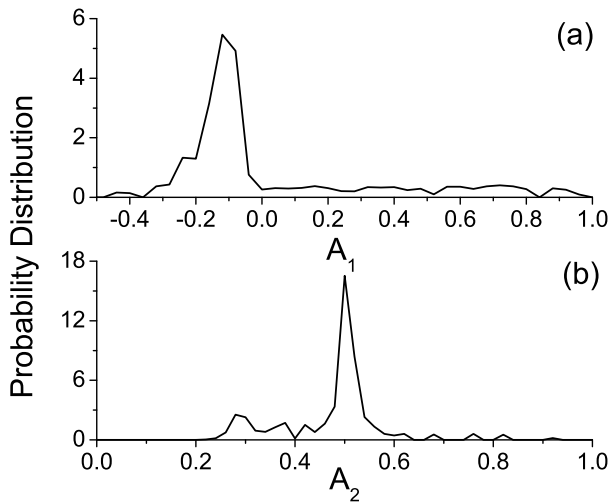}
\end{center}
\caption{Quantum distributions for prealigned $KCl$ molecules. Here the prealignment pulse ($P=25$) was parallel to the deflecting field (strong field, $1.8\cdot 10^7V/m$; $C=5.81\cdot
10^{-4}$ ; $D=9.62$). Temperature corresponds to $J_T=5$. These graphs are similar to the classical graphs of Fig. \ref{Classical_Parallel}.} \label{QuantumParallel}
\end{figure}

If the molecules are subject to a strong femtosecond pre-aligning
pulse parallel to the deflecting field, the corresponding
interaction potential is given, as in the previous section, by Eq. (\ref{U_general_eq}), in which
$F$ is replaced by the envelope $\epsilon$ of the femtosecond pulse (including the $1/2$ factor, as was explained in the previous section). If the pulse is short compared to the typical periods of molecular
rotation, it may be again considered as a delta-pulse. In the impulsive
approximation, one obtains the following relation between the
angular wavefunction before and after the pulse applied at $t=0$
\cite{Gershnabel}:
\begin{equation}
\Psi(t=0^+)=\exp(iP\cos^2\theta)\Psi(t=0^-),\label{before_after}
\end{equation}
where the kick strength, $P$ is given by Eq.(\ref{Kick_P}). For the
vertical polarization of the laser field, $m$ is a conserved quantum
number. This allows us to consider the excitation of the states with
different initial $m$ values separately. In order to find
$\Psi(t=0^+)$ for any initial state, we introduce an artificial
parameter $\xi$ that will be assigned the value $\xi=1$ at the end
of the calculations, and define
\begin{equation}
\Psi_{\xi}=\exp\left[(iP\cos^2\theta)\xi\right]\Psi(t=0^-)
=\sum_{J}c_J(\xi)|J,m\rangle.\label{xi_relation}
\end{equation}
By differentiating both sides of Eq.(\ref{xi_relation}) with respect
to $\xi$, we obtain the following set of differential equations for
the coefficients $c_J$:
\begin{equation}
\dot{c}_{J'}=iP\sum_J c_J\langle
J',m|\cos^2\theta|J,m\rangle,\label{Differential equations}
\end{equation}
where $\dot{c}= dc / d\xi $. The matrix elements in
Eq.(\ref{Differential equations}) can be found using recurrence
relations for the spherical harmonics \cite{Arfken}. Since
$\Psi_{\xi=0}=\Psi(t=0^-)$ and $\Psi_{\xi=1}=\Psi(t=0^+)$ (see
Eq.(\ref{xi_relation})), we solve numerically this set of equations
from $\xi=0$ to $\xi=1$, and find $\Psi(t=0^+)$. In order to
consider the effect of the field-free alignment at thermal
conditions, we repeated this procedure for every initial
$|J_0,m_0\rangle$ state. To find the modified population of the
$|J,m\rangle$ states, the corresponding contributions from different
initial states were summed together weighted with the Boltzmann's
statistical factors:
\begin{eqnarray}
f({\cal
A}_{J,m_0})&=&\sum_{J_0,\bar{J}}\frac{\exp(-E^{J_0}/k_B T)}{Q_{rot}}
\nonumber\\ &\times& |c_{\bar{J}}|^2\delta_{{\cal A}_{J,m_0},{\cal
A}_{\bar{J},m_0}},\label{QuantDistribution}
\end{eqnarray}
where $c_J$ are the coefficients (from Eq. \ref{Differential
equations}) of the wave packet that was excited from the initial
state $|J_0,m_0\rangle$; $\delta$ is the Kronecker delta symbol, and
$Q_{rot}$ is the rotational partition function. The distribution in
the case of parallel pre-alignment is given in Fig.
\ref{QuantumParallel}. The results are quite similar to the
classical results from Fig. \ref{Classical_Parallel}.

\begin{figure}[htb]
\begin{center}
\includegraphics[width=8cm]{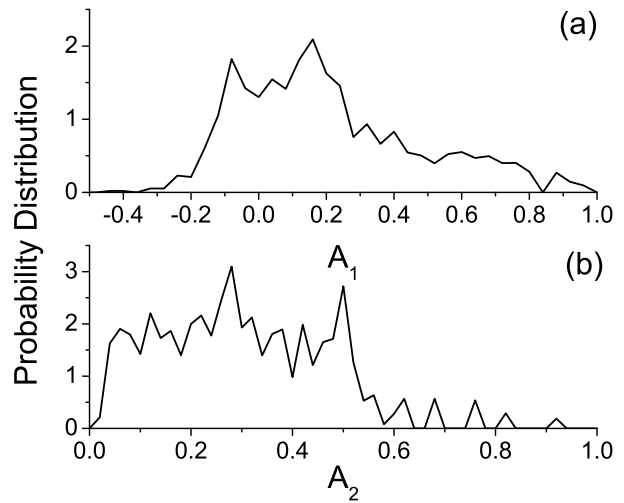}
\end{center}
\caption{Quantum distributions of $KCl$ molecules prealigned by the means of a pulse ($P=25$) perpendicular to the deflecting field (strong field: $1.8\cdot 10^7V/m$; $C=5.81\cdot 10^{-4}$ ; $D=9.62$). Temperature corresponds to $J_T=5$. These graphs are similar to the classical graphs of Fig. \ref{Classical_Perpendicular}.}
\label{QuantumPerpendicular}
\end{figure}

In the case of an aligning pulse in the $x$ direction, the operator
in Eq.(\ref{before_after}) becomes:
\begin{equation}
\Psi(t=0^+)=\exp(iP\cos^2\phi\sin^2\theta)\Psi(t=0^-),\label{before_after_perpendicular}
\end{equation}
and a  procedure similar to the described above is used to find the
deflection distribution (one should pay attention that $m$ is no
longer a conserved number during the operation of the pulse in the
$x$ direction). The distribution for  the case of perpendicular prealignment is given in Fig.
\ref{QuantumPerpendicular}.  These results are similar to the
classical predictions from Fig. \ref{Classical_Perpendicular}.

\section{Discussion and Conclusions} \label{Conclusions}

In this work we considered molecular deflection by weak and strong inhomogeneous static electric fields.  As the deflecting field is increased, it modifies the time-averaged alignment/orientation of the molecules. This affects the dipole force, and we have studied both classically and quantum mechanically the resulting deflection process. We found that laser induced field-free pre-alignment provides an effective tool for controlling molecular deflection. Depending on the polarization of prealignment pulse, different control actions may be exerted. In particaular, we predict a dramatic increase in the brightness of the scattered molecular beam, when the prealignment pulse is parallel to the direction of the deflecting field. Though we discussed (for simplicity) linear molecules in this work, a similar control mechanism may be considered for polyatomic molecules with more complicated geometry. Being in free space such molecules rotate about their own axis as well, which leads to ${\cal A}_2$ distribution different from that of Eq. (\ref{Gamma_Dist}), and ${\cal A}_1$ is not necessarily $0$. For such molecules the prealignment will play a significant role in reducing their dipole interaction with the static field. Molecular deflection by inhomogeneous static electric fields may be used for the  separation of molecular mixtures.  Narrowing  the distribution of the scattering angles may substantially increase the efficiency of separation of multi-component beams, especially when the pre-alignment is applied selectively to certain molecular species, such as isotopes \cite{isotopes}, or nuclear spin isomers
\cite{Fauchet,isomers}. Controlling molecular deflection by  means of laser-induced prealignment may be implemented also for magnetic molecules moving in the static inhomogeneous magnetic fields, and this phenomenon is a subject of the currently ongoing research. Controlling the dipole interaction by  laser-induced pre-alignment may find applications in  molecular deceleration methods using time and spatially varying electric and magnetic fields \cite{Deceleration}.

We acknowledge the support of our study  by a grant from the Israel Science Foundation. This research is made possible in part by the historic generosity of the
Harold Perlman Family. IA is an incumbent of the Patricia Elman Bildner Professorial Chair.

\section{Appendix: Calculation of ${\cal A}_{1,2}$ by  means of the theory of adiabatic invariants} \label{Appendix}

The energy of a molecule participating in the deflection process is:
\begin{equation}
H=\frac{1}{2}I\left( \dot{\theta}^2+\dot{\phi}^2\sin^2\theta
\right)-\mu F \cos\theta,\label{Energy_Equation}
\end{equation}
where we neglected the effect of polarizability (which is small for
the fields and molecules considered in the main body of the paper). The conjugate momenta
$P_{\phi}$ and $P_{\theta}$ are given by Eqs.(\ref{Pphi}) and
(\ref{Ptheta}), respectively, and $P_{\phi}$ is a constant of motion. It is convenient
to change variables and to define new constants
\cite{Dugourd,Goldstein}:
\begin{equation}
u\equiv -\cos\theta, \label{u_constant}
\end{equation}

\begin{equation}
\beta\equiv \frac{2}{I}H, \label{beta_constant}
\end{equation}

\begin{equation}
\alpha\equiv \frac{2\mu F}{I}.\label{alpha_constant}
\end{equation}

It is easy to show that $u$ obeys the following equation:
\begin{equation}
\left ( \frac{du}{dt} \right )^2=(\beta-\alpha u)(1-u^2) -
\left(\frac{P_{\phi}}{I}\right)^2,\label{g_definition}
\end{equation}
which gives:
\begin{equation}
dt=\frac{du}{\sqrt{g(u)}}, \label{dt_du_relation}
\end{equation}
where $g(u)$ is the rhs of the Eq.(\ref{g_definition}).
If $\alpha=0$,  the polynomial $g(u)$ has two roots,  $u_1,u_2$ such that
$-1\leq u_1\leq u_2 \leq 1$. When $\alpha\neq 0$,  $g(u)$ has three
roots (denoted by $u_1,u_2,u_3$). Let us  analyze the behavior of this polynomial.
If $u\gg 1$, then $g(u)\approx \alpha u^3>0$. If we substitute
$u=1$, then $g(1)=-\left(\frac{P_\phi}{I}\right)^2\leq 0$. That means
that the three real roots of $g(u)$ are ordered  as $-1\leq u_1\leq
u_2\leq1\leq u_3$. The polynomial $g(u)$ is positive between $u_1$ and $u_2$,
and each root corresponds to a real angle $\theta$. This allows
one to determine the half-period of the motion in the static field $F$ by integrating
Eq.(\ref{dt_du_relation}) from $u_1$ to $u_2$.

Since the potential is time-dependent (at least during the rising time),
the energy of the system is not a constant of motion. However, the
deflection potential is adiabatic with respect to the rotational
motion and, therefore, we can use adiabatic invariants to determine
the energy of the system \cite{Dugourd,Goldstein,Landau}. The
adiabatic invariant related to the coordinate $\theta$ is:
\begin{equation}
I_{\theta}=\int_{u_1}^{u_2} P_{\theta}d\theta.
\label{adiabatic_invariant}
\end{equation}
From Eqs.(\ref{Ptheta}), (\ref{g_definition}) and
(\ref{adiabatic_invariant}) it is easy to derive:
\begin{equation}
I_{\theta}=\kappa \int_{u_1}^{u_2} \frac{\sqrt{g(u)}}{1-u^2}du,
\label{adiabatic_invariant_modification}
\end{equation}
where $\kappa$ is a constant. The energy $H$ of the molecule as a
function of the energy $H_0$ without electric field is obtained
numerically by solving the following equation:
\begin{equation}
I_{\theta}=I_{\theta}^0,\label{numerical_E_adiabatic_invariant}
\end{equation}
where $I_{\theta}^0$ is calculated for $\alpha=0$, that is in the absence of the external field.

Once the energy of the system $H$ and the polynomial $g(u)$ are
found, ${\cal A}_{1,2}$ is given by:
\begin{equation}
{\cal A}_{1,2} =\frac{\int_{u_1}^{u_2}
(-u)^{1,2}du/\sqrt{g(u)}}{\int_{u_1}^{u_2} du/\sqrt{g(u)}}.
\label{Average_alignment_adiabatic_invariant}
\end{equation}
This is equivalent to calculating ${\cal A}_{1,2}$ with the help of
 Eq.(\ref{average_time_align}).

\bibliographystyle{phaip}

\begin{references}

\bibitem{McCarthy} T. J. McCarthy, M. T. Timko and D. R. Herschbach,
J. Chem. Phys. $\textbf{125}$, 133501 (2006).

\bibitem{Benichou} E. Benichou, A. R. Allouche, R. Antoine, M. Aubert-Frecon,
M. Bourgoin, M. Broyer, Ph. Dugourd, G. Hadinger and D. Rayane, Eur.
Phys. J. D. $\textbf{10}$, 233 (2000).

\bibitem{Loesch} H. J. Loesch, Chem. Phys. $\textbf{207}$, 427 (1996).

\bibitem{Antoine} R. Antoine, D. Rayane, A. R. Allouche, M. Aubert-Frecon, E. Benichou, F. W. Dalby, Ph. Dugourd,
M. Broyer and C. Guet, J. Chem. Phys. $\textbf{110}$, 5568 (1999).

\bibitem{Deflection_general} H. Stapelfeldt, H. Sakai, E. Constant and P. B. Corkum,
Phys. Rev. Lett. $\textbf{79}$, 2787 (1997); H. Sakai, A. Tarasevitch, J. Danilov, H. Stapelfeldt, R. W. Yip, C. Ellert, E. Constant and P. B. Corkum, Phys. Rev. A, $\textbf{57}$, 2794 (1998).


\bibitem{Lens} B. S. Zhao, H. S. Chung, K. Cho, S. H. Lee, S. Hwang, J. Yu, Y. H. Ahn, J. Y. Sohn, D. S. Kim, W. K. Kang, and D. S. Chung, Phys. Rev. Lett. $\textbf{85}$, 2705 (2000); H. S. Chung, B. S. Zhao, S. H. Lee, S. Hwang, K. Cho, S. H. Shim, S. M. Lim, W. K. Kang and D. S. Chung, J. Chem. Phys. $\textbf{114}$,
8293 (2001).

\bibitem{Prism} B. S. Zhao, S. H. Lee, H. S. Chung, S. Hwang, W. K. Kang, B. Friedrich and D. S. Chung, J. Chem. Phys. $\textbf{119}$,
8905 (2003).

\bibitem{Friedrich_ilya} B. Friedrich, Phys. Rev. A \textbf{61}, 025403 (2000).

\bibitem{Gordon_ilya} R. J. Gordon, L. Zhu, W. A. Schroeder and T. Seideman, J. Appl. Phys. \textbf{94}, 669 (2003).

\bibitem{Fulton_ilya} R. Fulton, A. I. Bishop and P. F. Barker, Phys. Rev. Lett. \textbf{93} 243004 (2004).

\bibitem {Bishop_ilya} R. Fulton, A. I. Bishop, M. N. Schneider, P. F. Barker, Nature Phys. \textbf{2}, 465 (2006).

\bibitem{Seideman} T. Seideman, J. Chem. Phys.  \textbf{106}, 2881 (1997);J. Chem. Phys. \textbf{107}, 10420
(1997); J. Chem. Phys. \textbf{111}, 4397 (1999).

\bibitem{Story} T. L Story and A. J. Hebert, J. Chem. Phys.
$\textbf{64}$, 855 (1976).

\bibitem{Schafer} R. Sch\"{a}fer, S. Schlecht, J. Woenckhaus and J. A. Becker, Phys. Rev. Lett. $\textbf{76}$, 471 (1996).

\bibitem{Brooks} P. R. Brooks, E. M. Jones and K. Smith, J. Chem. Phys.
$\textbf{51}$, 3073 (1969).

\bibitem{Kramer} K. H. Kramer and R. B. Bernstein, J. Chem. Phys.
$\textbf{42}$, 767 (1965).

\bibitem{Lubbert} A. L\"{u}bbert, G. Rotzoll and F. G\"{u}nther, J. Chem. Phys. $\textbf{69}$, 5174 (1978).

\bibitem{post} L. Holmegaard, J. H. Nielsen, I. Nevo and H. Stapelfeldt, Phys. Rev. Lett. $\textbf{102}$,
023001 (2009); F. Filsinger, J. K\"{u}pper, G. Meijer, L. Holmegaard, J. H. Nielsen, I. Nevo, J. L. Hansen and H. Stapelfeldt, J. Chem. Phys. $\textbf{131}$,
064309 (2009).

\bibitem{Zon} B. A. Zon and B. G. Katsnelson, Zh. Eksp. Teor. Fiz. \textbf{69},
1166 (1975) [Sov. Phys. JETP \textbf{42}, 595 (1975)].

\bibitem{Friedrich} B. Friedrich and D. Herschbach, Phys. Rev. Lett. $\textbf{74}$,
4623 (1995); J. Chem. Phys. $\textbf{111}$, 6157 (1999).

\bibitem{Barker-new} S.M. Purcell and P.F. Barker, Phys. Rev. Lett.
\textbf{103}, 153001 (2009).

\bibitem{Gershnabel_Deflection} E. Gershnabel and I. Sh. Averbukh, Phys. Rev. Lett. \textbf{104}, 153001 (2010); Phys. Rev. A \textbf{82}, 033401 (2010).

\bibitem{Stapelfeldt} H. Stapelfeldt and T. Seideman, Rev. Mod.
Phys. $\textbf{75}$, 543 (2003).

\bibitem{Stapelfeldt1} V. Kumarappan, S. S. Viftrup, L. Holmegaard, C. Z. Bisgaard and H. Stapelfeldt, Phys. Scr. $\textbf{76}$, C63 (2007).

\bibitem{3D1} J. J. Larsen, K. Hald, N. Bjerre and H. Stapelfeldt,  Phys. Rev. Lett. \textbf{85}, 2470 (2000).

\bibitem{3D2} J.G. Underwood, B. J. Sussman and A. Stolow, Phys. Rev.
Lett. \textbf{94}, 143002 (2005); K. F. Lee, D. M. Villeneuve, P. B. Corkum, A. Stolow and J. G. Underwood, Phys.
Rev. Lett. \textbf{97}, 173001 (2006).

\bibitem{3D3} S. S. Viftrup, V. Kumarappan, S. Trippe and H. Stapelfeldt, Phys. Rev. Lett. \textbf{99},
143602 (2007).

\bibitem{Tikhonov} G. Tikhonov, K. Wong, V. Kasperovich and V. V.
Kresin, Rev. Sc. Instr. $\textbf{73}$, 1204 (2002).

\bibitem{Ramsey} N.F. Ramsey, Molecular Beams (Oxford University Press, New York, 1956)


\bibitem{Gradshteyn} I. S. Gradshteyn and I. M. Ryzhik, $Table$ $of$ $Integrals$, $Series$,
$and$ $Products$, 7th ed. (Elsevier Academic Press, USA, 2007).

\bibitem{Gershnabel} E. Gershnabel, I. Sh. Averbukh and R. J. Gordon, Phys. Rev.
A, $\textbf{74}$, 053414 (2006).

\bibitem{Arfken} George B. Arfken, Hans J. Weber, $Mathematical$
$Methods$ $For$ $Physicists$, 6th ed. (Elsevier Academic Press, USA,
2005).

\bibitem{isotopes}  S. Fleischer, I. Sh. Averbukh and Y. Prior,  Phys. Rev. A $\textbf{74}$, 041403(R)
(2006).
\bibitem{Fauchet} M. Renard, E. Hertz, B. Lavorel, and  O. Faucher,  Phys. Rev. A $\textbf{69}$, 043401 (2004).
\bibitem{isomers}  S. Fleischer, I. Sh. Averbukh, and Y. Prior,  Phys. Rev. Lett., $\textbf{99}$, 093002
(2007); E. Gershnabel and I. Sh. Averbukh, Phys. Rev. A,
$\textbf{78}$, 063416 (2008).

\bibitem{Deceleration} J. A. Maddi, T. P. Dinneen and H. Gould,
Phys. Rev. A, $\textbf{60}$, 3882 (1999); E. Narevicius, C. G.
Parthey, A. Libson, M. F. Riedel, U. Even and M. G. Raizen, New J.
Phys., $\textbf{9}$, 96 (2007).

\bibitem{Dugourd}  P. Dugourd, I. Compagnon, F. Lepine, R. Antoine, D. Rayane and M. Broyer, Chem. Phys. Lett. $\textbf{336}$, 511 (2001).

\bibitem{Goldstein} H. Goldstein, C. Poole and J. Safko, $Classical$
$Mechanics$, 3rd ed. (Addison Wesley, USA, 2001).

\bibitem{Landau} L. D. Landau and E. M. Lifshitz, $Mechanics$, 3rd ed. (Butterworth-Heinemann, UK,
1976).



\end{references}

\end{document}